# Peculiarities of Ferro-Antiferroelectric Phase Transitions 9. Alternative for dipole-glass description of properties of substances with coexisting ferroelectric and antiferroelectric phases

V. M. Ishchuk[1] and V. L. Sobolev[2]

[1]*Science and Engineering Center "Reactivelektron", Donetsk, 83096 Ukraine,*

[2]*Department of Physics, South Dakota School of Mines and Technology, Rapid City, SD 57701*

**Abstract.**

It is demonstrated that the substances with small difference in the free energies of the ferroelectric and antiferroelectric phases possess a set of properties characteristic for the so-called "dipole glasses". Possible phase diagrams of the substances that can be misguidedly attributed to glasses are discussed. Main attention has been paid to the process of long-time relaxation of physical characteristics of these compounds after their state of thermodynamic equilibrium was disturbed by external influences. The long-time relaxation along with the pronounced frequency dependence of parameters (for example, dependence of dielectric or magnetic characteristics on the frequency of measuring field) is considered as main features due to which these systems are classified as dipole glasses. Our main purpose is to call attention to the fact that one has to be cautious during the interpretation of experimental results in substances with inhomogeneous states of coexisting domains of ferroelectric and antiferroelectric phases.

PACS numbers: 77.84.-s, 77.80.Dj, 77.80.B-, 77.80.bn

## 1. Introduction

The concept - "spin glass" - was introduced in physics of magnetic phenomena in the mid seventies of the last century [1-3]. Soon the notion of "dipole glass" was also suggested by analogy with magnetism. It has been widely used in discussions of experimental results for the so-called "relaxor" ferroelectric and KDP-type substances with order-disorder phase transitions (PT).

The substances, which are usually attributed to "dipole glasses", possess the following main properties:

- an essential dispersion of dielectric permittivity in the region of diffuse maximum on the dependences $\varepsilon'(T)$ and $\varepsilon''(T)$ with fulfilment of the Vogel-Fulcher law: $\omega = (1/\tau_0)\exp\{-\Delta/k(T_m - T_f)\}$, where $\Delta$ and $\tau_0$ are material parameters, $k$ is the Boltzmann constant, $T_m$ is the temperature of $\varepsilon'(T)$ and $\varepsilon''(T)$ maximum ($T'_m$ or $T''_m$), $T_f$ is an effective temperature different for $(T'_m$ and $T''_m)$;

- an increase of the frequency of the AC field causes decrease of the maximum value of the real component of permittivity, whereas for the imaginary component this magnitude increases, the temperature of $\varepsilon'(T)$ maximum being always higher than that of $\varepsilon''(T)$ maximum;

- an increase of the amplitude $E_0$ of the measuring field leads to linear decrease of $T'_m$ and to increase of $\varepsilon'(T'_m)$; the tangent of the $T'_m(E_0)$ dependence decreases with increasing AC field frequency;

- hysteresis loops of these substances have a specific form: narrow dielectric hysteresis loops with a small residual polarization and narrow quadratic loops of electrooptic hysteresis;

- a presence of effects which point to the existence of polar phase micro/nano domains at temperatures essentially higher than $T'_m$ and $T''_m$;

- dependence of properties of these substances on the sample's history;

- long-duration relaxation;

- high degree of diffuseness of the paraelectric phase transition.

The above-listed properties may be somewhat varied, or may manifest themselves not in the complete set. As a rule, the fulfilment of the first, sixth and seventh condition is dominating.

It was demonstrated [4-9] that the above set of properties is typical for the substances in which ferroelectric (FE) and/or antiferroelectric (AFE) ordering may be present, and the difference in free energies of these states is small. Phase transitions between ferroelectric and antiferroelectric phases (FE-AFE PT) may take place in these substances under the action of external factors such as temperature, field or pressure.

In this paper, we will focus our main attention on the process of the long-time relaxation of properties and physical characteristics of these substances after their state of thermodynamic equilibrium was disturbed by external influences. Such long-time relaxation along with the pronounced frequency dependence of parameters (for example, dependence of dielectric or magnetic characteristics on the frequency of measuring field) is considered as a feature of the spin or dipole glass. The substances that manifest such properties are referred to as "glasses" almost immediately.

We will demonstrate that the above-mentioned relaxation is also manifested in the substances in which FE and AFE phases are realized and the difference between their free energies is small. After that, we will discuss possible phase diagrams of substances that can mimic themselves as "glasses".

By no means have we wanted to cast doubt on the existence of "dipole glasses" in the nature. Our main purpose is to call attention to the fact that one has to be careful during the interpretation of experimental results in substances in which the inhomogeneous states with coexisting domains of FE and AFE phases can take place. The peculiarities of behaviour of such substances can be really misleading during interpretation. More over the existence of such states (stable in the absence of external influences) was considered impossible in a wide class of substances until recently.

**2. Model Concepts**

The behaviour of the system that has a small difference in the free energies of the FE and AFE phases and domains of these phases may coexist in the sample was considered in [6]. It was shown that under the action of DC electric field the state of the substance changes due to the displacement of interphase domain walls (IDW) while the internal state within the domains remains unchanged. Such motion of IDW is inertial and accompanied by relaxation processes. The relaxation dynamics of IDW should manifest itself most vividly when the said substances are subjected to the action of the AC electric field.

The dynamics of IDW under the action of the AC electric field can be examined using the equation describing the balance of forces acting on IDW. This equation representing the condition of balance of forces has the form:

$$P_{E(t)} + R(u) + P(u,t) = 0. \qquad (1)$$

In this equation $u(t)$ is the displacement of the IDW, $E(t)$ is the electric field intensity, $P_{E(t)}$ is the pressure acting on IDW due to the expansion of FE domain volume in external field. The forces which counteract the IDW displacement under the effect of the field can be presented as a sum of two terms: here $R(u)$ is the pressure of the forces caused by the interaction of the IDW with immobile crystal structure defects, and $P(u, t)$ is the pressure of the so-called after-effect forces. The latter forces may be caused by different factors: the interaction of the IDW with mobile crystal lattice defects, as well as nonzero duration of the thermal processes connected with phase transitions in those regions of the crystal through which IDW passes.

Equation (1) describes the equilibrium of the IDW under the action of different kinds of physical forces. While considering this relation as a motion equation (Newton equation), one can easily see that the terms $m_W \ddot{u}(t)$ ($m$ is the effective mass of the IDW) and $\gamma m_W \dot{u}(t)$ are not taken into account. These terms may not be considered during the analysis of motion if the effective mass $m_W$ is small and if the frequency of the measuring AC electric field is essentially lower than the frequency of the IDW oscillation in the potential well $\int du R(u)$ created by immobile crystal lattice defects. Under above-mentioned assumptions Eq.1 has the form of the integral equation:

$$[1+\eta G(t)]u(t) - \eta \exp(-t/\tau)\int_0^t u(t')\exp(t'/\tau)\frac{dt'}{\tau} = u_0 \exp(i\omega t) \qquad (2)$$

where $\eta$ is the substance's constant, $u_0 = PE_0/\alpha$ is the IDW displacement under the action of DC electric field of the intensity $E_0$, $\alpha$ is the elastic constant defined by the equation $R(u) = -\alpha u$, $(\alpha > 0)$ and $G(t) = 1 - \exp(-t/\tau)$ is the time function. For considered features, relaxation time $\tau$ is defined by the equation:

$$\tau = \tau_o \exp(\Delta/kT). \qquad (2.a)$$

Here $\Delta$ and $\tau_0$ are the same material parameters as in the Vogel-Fulcher formula. The following solution of Eq. 2 can be obtained [6] under the condition $G = 1$ (that is valid for $t \to \infty$)

$$\chi = \chi' - i\chi'' = \chi_0 \frac{1+(\omega\tau)^2(1+\eta)-i(\omega\tau)\eta}{1+(\omega\tau)^2(1+\eta)^2}. \qquad (3)$$

Susceptibility $\chi(t)$ is given by the formula $\chi(t) = P_s Su(t)/E(t)$, where $S$ is the area of the oscillating IDW, $P_s$ is spontaneous polarization, and $\chi_0 = P_s^2 S/\alpha$ is the static susceptibility. Thus,

$$\chi'(\omega,T) = \chi_0 \frac{1+(\omega\tau)^2(1+\eta)}{1+(\omega\tau)^2(1+\eta)^2} \equiv \chi_0(T)F_1(\omega,T), \qquad (4a)$$

$$\chi''(\omega,T) = \chi_0 \frac{\eta(\omega\tau)}{1+(\omega\tau)^2(1+\eta)^2} \equiv \chi_0(T)F_2(\omega,T). \qquad (4b)$$

Based on Eq.4 one can obtain equation that relates the position of maximum of the temperature dependence $\varepsilon'(T)$ and the frequency of the measuring field [6]

$$\frac{1}{T'_m - T_f} = -\frac{k}{\Delta}\ln\omega - \frac{k}{\Delta}\ln\tau_0.$$

The above relation can be also written in the Vogel-Fulcher form:

$$\omega = (1/\tau_0)\cdot\exp\{-\Delta/k(T'_m - T_f)\}.$$

The interphase domain wall separates the domains with FE and AFE states, for which elementary crystal cells have different size. For most of known ferroelectrics and antiferroelectrics such difference is essential (the review of this problem see, e.g. [10]). The method of transmission electron microscopy was used to investigate the structure of domains of the coexisting FE and AFE phases [10-12]. It was shown that IDW under consideration had a coherent character. This signifies that the crossing of the IDW (from one phase to the other) is accompanied with continuous conjugation of the atom planes (free of breaks and dislocations). Such coherent structure of the IDW leads to an increase of elastic energy in the vicinity of IDW.

This increase is the more essential the larger is the difference in the configuration volumes of the FE and AFE phases.

In the substances under consideration equivalent crystallographic sites are occupied by ions which differ either in size or in charge or in both of them. In a single-phase state (or inside domains of each of the coexisting phases), in the absence of external factors each of the lattice-forming ions is not subjected to the action of forces (more correctly, the resultant force which affects each ion is equal to zero). Entirely opposite situation takes place for the ions located near the "bare" IDB. The balance of forces affecting each of the ions is upset. "Large" ions are pushed out into those domains which have a larger configuration volume and, consequently, a larger distance between crystal planes. "Small" ions are shifted into the domains with a smaller configuration volume and a smaller interplanar distance. Such process brings about the decrease of the elastic energy concentrated along the "bare" IDW and the increase of the energy bound up with the segregation of the substance. The considered process of ion segregation will be completed when the structure of the newly-formed IDW provides the energy minimum. Such a "clothed" IDW will be further called a real IDW or simply IDW. If equivalent sites of the crystal lattice are occupied by ions which have not only different size but also different charges, then the chemical segregation within the IDB limits is accompanied by the local violation of electroneutrality.

We have shown [8, 9, 13] that the formation of the heterophase structure of coexisting FE and AFE domains is accompanied by the emergence of chemical inhomogeneity of the substance. Here we are to strictly emphasize that it is just phase inhomogeneity (coexistence of the FE and AFE phases) that leads to chemical segregations, but not on the contrary.

The process of chemical segregation in the vicinity of the FE-AFE phase boundary can take place only at those conditions that allow the existence and coexistence of the dipole ordered FE and AFE phases. At high temperatures (when the uniform paraelectric state is present in the substance) driving forces for chemical segregation are absent, and the process of the substance annealing leads to the chemical homogeneity. Thus, the discussed above process of chemical segregation in the vicinity of the FE-AFE phase boundary is reversible in the course of temperature cycling during cooling and heating.

If the IDW is displaced under the action of electric field (or appears after cooling from paraelectric state), then the process of chemical segregation will occur in the vicinity of the position of this new IDW. "Old" chemical segregations will be cleaned out. However, it is not only the IDW itself that controls the process of chemical segregation in the vicinity of its location. There exists an influence opposite to the chemical segregation on the IDW during its motion (see Eq. (1)). The said segregation becomes a mobile defect of the crystal lattice. The interaction between segregation and IDW should be the most pronounced when the formation of such segregation is accompanied by a violation of local electroneutrality.

The mentioned interrelation between the IDW and the formation of mobile defects of the crystal lattice lead to peculiarities in the IDW dynamics. At the temperatures, at which FE and AFE states are realized, the rate of ionic diffusion is low. Equation (2) contains the time-function $G(t) = 1 - \exp(-t/\tau)$. We adopted condition $G = 1$ (for $t \to \infty$). Now in view of the fact that the above mentioned process of chemical segregation is the long-duration one, the time-dependent function $G(t)$ has to be substituted for $G = 1$ in (2), since the condition $t \to \infty$ is not fulfilled in real experiment. In this case more complicated expressions for $\varepsilon'$ and $\varepsilon''$ can be obtained instead (4):

$$\varepsilon'(\omega,T) = \varepsilon_0 \frac{1+(\omega\tau)^2[1+\eta G(t)]}{1+(\omega\tau)^2[1+\eta G(t)]^2}, \qquad (5a)$$

$$\varepsilon''(\omega,T) = \varepsilon_0 \frac{(\omega\tau)\eta G(t)}{1+(\omega\tau)^2[1+\eta G(t)]^2}. \qquad (5b)$$

Under the measurement's conditions (when the period of oscillations of the measuring field is much shorter than characteristic time of diffusion processes) $\omega\tau \to \infty$ we have:

$$\frac{\varepsilon'}{\varepsilon_0} = \frac{1}{1+\eta G(t)}. \qquad (6)$$

As one can see from this expression, the dielectric constant reaches its equilibrium value during a long period of time after the action of any factor leading to the shift the IDWs (or after their appearance during cooling from high temperatures) because at the temperatures $T < T_c$ the coefficients of diffusion are small. The time dependence of $(\varepsilon'/\varepsilon_0)$ for some values of τ is shown in Fig.1a for particular case of η = 1.

Let us remind (see eq. 2a) that the relaxation time τ is determined by the activation energy Δ for the processes under consideration. The extended in time scale of the process of change of the chemical composition near the "bare" IDW takes place in the course of formation of segregates. It is naturally that the change of the activation energy (Δ = Δ(t)) takes place along with the segregates formation process. Therefore, the characteristic relaxation time also depends on time (τ = τ(t)). A small number of experiments on establishment of dependency τ(t) has been done up to now. We will use the results of investigation of the long-time relaxation [8, 9]. The process of formation of segregates along the IDWs takes place during more than 30 hours.

Experimental studies were carried out using two series of PZT-based solid solutions, namely, $Pb_{0.85}(Li_{1/2}La_{1/2})_{0.15}(Zr_{1-y}Ti_y)O_3$ (15/100-Y/Y PLLZT) and $Pb_{1-3x/3.91}La_x(Zr_{1-y}Ti_y)O_3$ with

x = 0.06 (6/100-Y/Y PLZT). Dependencies of crystal lattice parameters on Ti-concentration are shown in Fig.2 for both series of solid solutions. As one can see the solid solutions in which coexistence of the Fe (Rh) and the AFE (T) phases in the sample volume can be found in both cases. Investigations of the dynamics of formation (growth) of segregates were carried out in the following way. Samples were annealed at the temperatures (600-650) $^oC$ during 22 hours. At these temperatures the phase state of both series of solid solutions is a single phase paraelectric state. Then the samples were quenched down to room temperature and X-ray diffraction patterns by means of the Debye-Scherrer method were recorded in wide interval of angles after particular time intervals (details of experiments and obtained results one can find in [9, 10]). The inhomogeneous state of coexisting domains of the FE and AFE phases is formed in the samples after their quenching down to room temperature because the Curie temperatures of the investigated solid solutions were of the order of (170-180) $^oC$. The local decomposition (and formation of segregates) in the vicinity of the domain boundaries takes place. The diffuse X-ray lines with intensity much lower than the one of Bragg lines appear on the X-ray diffractograms due to the process of formation and growth of segregates in the vicinity of the FE-AFE domain boundaries. The dynamics of segregates formation determines the changes in intensity, position, and shape of these diffuse lines.

Time dependencies of intensities of diffuse X-ray lines for the 15/77/23 PLLZT and the 6/73/27 PLZT solid solutions for this time interval are presented in Fig.3 in logarithmic scale along both axes. As one can see in this figure the linear dependence is observed with the high level of accuracy. This clearly demonstrates that the time dependence of the relaxation time has a power character $\tau(t) = At^n$ with $n < 1$. Hence the time dependence of the dielectric permittivity in the process of aging (in the course of formation of segregates in the process of their growth) is given by the formula

$$\frac{\varepsilon'}{\varepsilon_0} = \frac{1}{1+\eta\left[1-\exp(-t^{1-n}/A)\right]}, (n<1). \tag{7}$$

The dependencies described by the formula (7) are presented in Fig. 1b.

Relaxation curves typical for spin glass systems are presented in Fig.1c and Fig.1d. These curves were obtained during the aging experiments measured after the switching of the external field and after heating up samples (the data is taken from [16]). As one can see, the dependence (7) completely describes the time behaviour of the systems, which are referred to as glasses in the process of the aging. It has to be stressed that the local decomposition of the solid solution caused by the local mechanical stresses in the vicinity of IDWs was the physical factor that determined the shape of the dependency (7).

## 3. EXPERIMENTAL RESULTS.

3.1. *Glass-like behaviour of perovskite oxides with coexisting of FE and AFE phases*

The aging of PLZT ceramic samples with composition 9.5/65/35 after quenching from high temperatures was investigated in [17]. We selected this study as an example due to the following reason. It was unambiguously shown that in the samples analogous in composition and location in the phase diagram (see [10,12,14]) the local decomposition of the solid solution and the formation of segregates in the vicinity of the IDWs separating domains of the FE and AFE phases takes place after quenching from high temperatures. As one can see in Fig.4 the real aging process may well be described by means of the formula (7). More over we have to ascertain that the expression (7) can describe even more protracted relaxation processes.

Phenomena of the so-called thermal dielectric, field, or mechanical memory that were experimentally observed in "relaxor ferroelectrics" of the PLZT or the PMN type are even more interesting from the physical point of view. All above mentioned effects have common physical basis, namely, the local decomposition of the solid solution in the vicinity of IDWs. For that reason, we will discuss only one effect – the phenomenon of thermal dielectric memory – in detail. This effect implies that when the samples are cooled to some particular temperature $T = T_{age}$ after annealing at high temperatures and are subjected to aging (during the time interval of 20 hours and longer) at this temperature a specific characteristic feature is observed on the $\varepsilon(T)$ dependencies at following temperature cycling. This feature consist in a "drop-shaped" behaviour of the dependence $\varepsilon(T)$ in the vicinity of the aging temperature $T_{age}$. Typical results of experiments [17, 18] are presented in Fig.5.

These results can have simple unambiguous explanation if one takes into account the inhomogeneous structure of coexisting domains of the FE and AFE phases present in the sample. Domain structure is formed in the bulk of the sample after the high temperature annealing. The share of each phase (and consequently the sizes of domains) is determined by the relation between the free energies of these phases. Diffuse local decomposition of the solid solution constantly takes place in the vicinity of the interdomain boundaries. However, this is the long time process and it is practically not manifested in material parameters. This influence is extremely small due to the fact that the characteristic time of measurement (during the temperautre measurements), which is of the order of several degrees Celcius per minute, is much shorter than the characteristic relaxation times of diffusion processes (which are of the order of tens of hours). The same is true for the external field or mechanical stress measurements. During aging at the temperature $T_{age}$, which lasts for long time (tens of hours), the local decomposition of the solid solution takes place and the structure of the formed segregates repeats the structure of the interphase boundaries. The longer is the the aging time the more proniounced is the spatial structure (the distinctive network) of segregates and the larger vloume of the sample is occupied by segregates. The dielectric permittivity is reduced as a result (see Fig.5). The formed spatial structure of segregates is conserved during the temperature cycling (with the several degrees per minute rate of the temperature change) following the aging.

The equilibrium relation between the shares of each phase and the sizes of domains of the FE and AFE phases are changing in the process of temperature cycling. The sizes and the structure of domains coincide with the preserved spatial structure of segregates when the temperature $T_{age}$ is achieved. At this moment the effective pining of interphase boundaries takes place and the contribution of oscillations of interdomain boundaries which determine the main contribution to the dielectric permittivity decreases. This pining of the interdomain boundaries leads to the appearance of the "drop-shaped" feature on the dependence ε($T$) (see Fig 5).

The nature of the effect of dielectric memory at the temperatures both above $T_m$ and below $T_m$ is the same as it clearly seen from the results of [18] (see also [22] and articles that appeared later). However, the authors of numerous papers who were trying to connect this phenomenon with relaxor ferroelectrics were forced to come up with different mechanisms for these two cases. One does not need to do that using the model suggested here. As shown in [7, 15] the structure of coexisting domains of the FE and AFE phases exists both above $T_m$ and below $T_m$. The X-ray studies carried out above $T_m$ demonstrated that the coexisting domains of the FE and AFE phases constitute the two-phase FE+AFE domain. This topic will be discussed in greater details in the next section of this article. Thus, the IDW structure that determines the effect of the dielectric memory exists both below $T_m$ and above $T_m$.

The shares of the AFE and FE phases and, thus, the spatial structure of IDWs can be also changed by means of other thermodynamic parameters such as external field and mechanical stress. The long time aging of samples at some nonzero values of above mentioned parameters will lead to formation of the new spatial structure of segregates and the effects of the filed or mechanical dielectric memory will be manifested during the cyclic changes of these parameters after the aging [19-21]. These effects are presented in Fig.6 and Fig. 7.

The effects discussed above are as a rule attributed to the so-called dipole-glass-like behaviour in relaxor systems. However, based on this dipole-glass approach it is impossible to describe and explain all manifestation of other phenomena (see the next section of this article). We demonstrated that rather simple explanation can be given if one takes into account coexistence of the AFE and FE phases.

We have considered above the effect of dielectric memory using the PLZT family of solid solutions as example. Similar effect is also characteristic of PMN based solid solutions. The result obtained for single crystals with composition $(1–x)Pb[(Mg_{1/3}Nb_{2/3})O_3 – xPbTiO_3$ with x = 0.10 and 0.12 one can find in [22] (phase diagram in presented in Fig.8). They are analogous to the result for PLZT given above. Some examples for different aging temperatures (both above and below $T_m$) and different Ti-content are shown in Fig.8. The physical mechanism of the effect is the same as in PLZT.

3.2. *Peculiarities of "composition–temperature" phase diagrams of order-disorder systems*

We have demonstrated that taking into account the coexistence of the FE and AFE phases one can give consistent explanation of all effects, which were attempted to attribute to the dipole-glass state in the so-called relaxor ferroelectrics. In what follows, we will try to show how this approach can be slightly extended.

Let us consider experimental "composition – temperature" phase diagrams of the solid solutions in which one of the components is a ferroelectric and the other is an antiferroelectric. The phase diagrams of the solid solutions for which FE and AFE nature of the low-temperature (LT) state is unambiguously identified are shown in Fig. 9 [23-26]. Phase diagrams for oxides with perovskite structure are presented in this figure. The presence of a sharp drop on the dependences $T_c(x)$ observed for the solid solutions located near the boundary which separates the regions with FE and AFE ordering is a common feature of all the presented phase diagrams. Such drop in the value of $T_c$ may reach 200° and more. It should be also noted that for all the solid solutions which phase diagrams are shown in Fig.9 the diffuseness of PE PT noticeably increases as the solid solution composition approaches the FE-AFE phase boundary in the diagram. As shown in ref. 4, 7, 15, such behaviour is caused by the existence of two-phase (FE+AFE) domains in the PE matrix of the substance at $T > T_c$ and the presence of ion segregations in the vicinity of the interdomain boundaries (see also Fig.11).

The symmetry and properties of the high temperature (HT) phase define all possible low temperature phases and PT under any temperature changes. In Fig.10 the typical temperature dependence is shown for the inverse dielectric constant (curve 1) at diffusive phase transitions (DPT) in the vicinity of FE-AFE-PE triple point. Thereat a special attention has to be paid to the value of the Curie-Weiss temperature $T_{cw}$, since it is above the Curie point (to the right of $\varepsilon$ (T) dependence maximum). From the viewpoint of the physics of phase transitions it is just at $T_{cw}$ point at which the transition into ordered state must be realised. Curve 3 built in agreement with the Curie-Weiss law presents the temperature dependence of $\varepsilon$ which reflects the properties of the high temperature phase and, thus, must describe the behaviour of dielectric constant when approaching the Curie point from above. Nevertheless, in the vicinity of the transition the experimental curves ε(T) and, consequently, 1/ε(T) are shifted towards lower temperatures (curves 1 and 2). In other words, at lowering the temperature the PT is blocked, displaced to the region of lower temperatures. Moreover, due to the diffuseness of the transition the said blocking should differ in intensity in different local regions of the crystal.

In our opinion, the main difficulty in understanding of physics of PT in the vicinity of triple point is to clarify the mechanism (or mechanisms) of "pressing out" the ordered phase towards the low-temperature region. Our results [4, 7, 15] allow to propose a mechanism of "pressing out" paraelectric PT towards the low-temperature region with its simultaneous diffuseness. The decisive factor which defines the said PT behaviour is the above-considered mechanism of ions separation at the interphase FE-AFE boundary into the two-phase domains.

The dynamics of DPT may be presented as follows. At the first step, in the process of cooling from high temperatures, the behaviour of such substance corresponds to the classical theory of phase transformations. The temperature dependence of dielectric constant obeys the Curie-Weiss law. Then the discussed two-phase FE+AFE domains appear at cooling. This fact is confirmed by the results of X-ray studies [15]. At the interphase FE-AFE boundary, separation of the elements forming the crystal lattice of the given substance (and segregations birth) is initiated due to the difference in their ionic radii. There appear local segregations with "foreign" composition and mesoscopic dimensions; in the general case they are not FE-active or AFE-active. It is just at this step (at temperature $T_{seg}$) that the deviation from the Curie-Weiss law manifests itself on 1/ε(T) and, consequently, ε(T) dependences. In the process of cooling the share of such segregations in the volume of the substance increases, and the deviation from the Curie-Weiss law becomes more noticeable.

However, passive dielectric segregations reduce both the share of the FE component in the substance and the value of dielectric constant. It is necessary to note that the influence of passive dielectric segregations is not limited to the above mentioned reduction of the FE phase share and dielectric constant. Resent achievements in physics of inhomogeneous solids testify that the introduction of non-magnetic impurities into magnetic substances or of passive dielectric ones into ferroelectrics and antiferroelectrics suppresses the formation of low temperature ordered phase. The literature on this problem is ample enough, but here we would like only to dwell on the well-investigated case of percolation systems. It is known at present [27] that when approaching the percolation threshold from the ordered phase (this case corresponds to the increase of the content of passive impurities in the system), the temperature of PT decreases and tends to zero at $x \to x_c$. Similar mechanisms work in the systems under discussion. The segregations which are developing in the process of cooling are passive dielectrics, so they "press out" the low temperature phase and, consequently, PT, towards lower temperatures. At present this problem has been investigated in detail not only for point defects of crystal lattice, but also for extended (mesoscopic-scale) inhomogeneities [28].

The above-said is valid for AFE PT, too. The distinction from FE PT consists in the following details. The transition from PE into AFE state is caused by condensation of non-polar mode with the wave vector at the boundary of the Brillouin zone, whereas the increase of dielectric constant at $T \to T_c$ results from softening (but not from condensation) of the polar mode with the wave vector in the center of the Brillouin zone. The latter mode interacts with soft non-polar mode. Therefore, in the case of AFE extrapolation of the straight-line part of the dependence $1/\varepsilon(T)$ does not yield the point of AFE PT (i.e. the point of stability loss for the high-symmetry phase with respect to relatively small anti-polar displacements of the crystal lattice ions).

Thus, in the proposed model of PT "pressing out", the basic role belongs to chemical segregations caused by the difference in configuration volumes of the FE and AFE phases. Such segregations were studied in [8, 9, 13].

Diffuseness of the PE PT essentially depends on the type of low temperature state as it is shown in the upper part of Fig.11 (the physical nature of the diffuseness parameter $\delta$ is considered in greater details in [7, 10], for example). For the regions of FE or AFE states remote from the triple point, $\delta$ is small. As the triple point is approached by changing the external parameters P (pressure) or $x$ (solid solution composition), the diffuseness of PT increases (see, for instance, [7, 10, 29]).

In all above-considered cases of oxides with perovskite structure (Fig. 9) the solid solution components had rather high proper values of the Curie point. Due to this, the minima on the dependence $T_c(x)$ do not reach zero of the Kelvin temperature scale. The phase diagrams of the solid solution with components having low values of the Curie temperatures are schematically shown in Fig.12 (at the top). The dependence of the segregation temperature on composition $T_{seg}(x)$ is denoted by dotted line. We would remind that $T_{seg}$ is the temperature at which the two-phase FE+AFE domains in PE matrix of the solid solution start to appear. At the same time at this temperature, the segregates start to emerge when temperature of the sample is decreasing from high temperatures. The bell-like dependence of the diffuseness parameter $\delta(x)$ (shown in Fig.11 at the top) is typical for all discussed solid solutions and promotes such behaviour of $T_{seg}(x)$ dependence.

To conclude the consideration of the diagram shown at the top of Fig.12, we would like to discuss possible nature of the line $T_{quant}(x)$ in the phase diagram. In our opinion, the nature is clear. A lot of attention is has been paid lately to $SrTiO_3$ and $SrTiO_3$-based solid solutions as an example of quantum ferroelectrics. The Curie point of these solutions lowers with the increase of content of the strontium titanate. However, in these solid solutions the PE phase transitions typical for ordinary ferroelectrics do not take place at changing the temperature. When the dependence $T_c(x)$ reaches the region of about 20-30 $K$ further decrease of the Curie point is not observed, and the character of the dependence $\varepsilon(E)$ changes. Such behaviour is explained by the quantum effects contribution which determines the nature of the line $T_{quant}(x)$ in the phase diagram presented at the top of the Fig. 12.

The majority of studies devoted to investigation of dipole-glass state is performed on series of solid solutions $K_{1-x}(NH_4)_xH_2PO_4$, $Rb_{1-x}(NH_4)_xH_2PO_4$, $Rb_{1-x}(ND_4)_xD_2AsO_4$ that fall into the class order-disorder ferroelectrics. Let us remind that the first component of these solid solutions is a ferroelectric and the second is an antiferroelectric. Experimental phase diagrams of $K_{1-x}(NH_4)_xH_2PO_4$, $Rb_{1-x}(NH_4)_xH_2PO_4$, and $Rb_{1-x}(ND_4)_xD_2AsO_4$ solid solutions are presented in Fig.12 . These diagrams were built on the base of the results of measurements carried out by different methods and presented in [30, 33, 35]. One can clearly see that there is practically complete coincidence of these experimental diagrams with the model one (presented on the top of the Fig.12). However, the latter is typical for those substances in which the phases with FE and AFE ordering are realized and domains of these phases coexist, while the experimental diagrams belong to the substances are classified nowadays as "dipole glasses".

For this reason, it seems expedient to analyze some experimental results obtained during investigations of phase transitions in the $K_{1-x}(NH_4)_xH_2PO_4$ and $Rb_{1-x}(NH_4)_xH_2PO_4$ systems of solid solution (and other related substances). We are going to consider the results that could not find their explanations in the scope of traditional approach and can be explained from the viewpoint of the effects caused by the coexisting FE and AFE phases. We would like to emphasize that in most cases such experimental results are consistent with the recently developed ideas about FE-AFE phase transformations with taking into account the interactions between coexisting phases (interaction between the domains of these phases). Moreover, there exist some results, which interpretation from another viewpoint seems to be artificial.

In this respect, we would like to note the experimental results [36] of studies of the diffuseness of the PE transition in $K_{1-x}(NH_4)_xH_2PO_4$ ($0 \leq x \leq 0.24$) solid solutions. These results are presented in Fig.13. The dependencies $\varepsilon(T)$ for solid solutions with $x$ от 0 до 0.24 are presented here (let us remind that the phase boundary betwenn the "dipole-glass" state and FE state (Fig.12) is at x = 0.20). As one can see the diffuseness parameter $\delta(x)$ increases with increase of $x$ and does not have any peculiarities at x = 0.20. There is an interesting detail clearly seen in Fig.12. The temperatures at which the deviation form the Curie-Weiss law starts ($T_{seg}(x)$ in our notations) are very close to the line $T_m(x)$ in the phase diagram for $K_{1-x}(NH_4)_xH_2PO_4$ in Fig. 12. Both these experimantal results are in complete agreement with the concept about coexisting domains of the AFE and FE phses that were presented above and discussed in the course of this article. We would like to call ones attention to the fact that the character of depencdencies $\varepsilon(T)$ does not change in any way when the composition of the solid solution crosses the phase boundary x = 0.20 in the phase diagram of the $K_{1-x}(NH_4)_xH_2PO_4$ solid solutions. It is apparent that the mechanism of diffuseness of the phase transition for the solid solutions with x < 0.20 and x > 0.20 is the same.

The results of studies of diffuseness of the phase transition in the presence of DC electric field applied to samples are even more interesting. The degree of the diffuseness parameter for the $K_{0.76}(NH_4)_{0.24}PO_4$ solid solution reduces as the field intensity increases [36] (Fig.14). In the process of the ordinary FE-PE phase transformations electric field leads to increase of the degree of transition diffuseness. However, in those substances for which FE and AFE orderings have a small difference in their energies and the FE and AFE phase domains coexist in the bulk of the samples, the described behaviour is normal. The stability of the FE state increases as against that of the AFE state when the field intensity increases. This is equivalent to shift of the position of the sample's state out of the FE-AFE-PE triple point in the phase diagram of states (from point 2 to point 1 in Fig.11). As seen from the upper part of this figure, the parameter of PT diffuseness decreases. This fact was revealed in experiments on $K_{0.76}(NH_4)_{0.24}PO_4$.

The described above experiments can be unambiguously explained if one takes into account the coexistence of the domains of the AFE and FE phases in those solid solutions in which one of the components is a ferroelectric and the other is an antiferroelectrics. Two circumstances lead to the rise of stability of the FE phase relatively to the AFE phase. First is the increase of the intensity of electric field and the second is the decrease of content $x$. The reduction of the share of the AFE phase in the sample volume lead to the reduction of the diffuseness of the phase transition (see [7, 10, 29] for details).

Presented results point to the essential influence of the coexistence of phases on the kinetics of the phase transition in order-disorder type substances in which one component is a ferroelectric and the other component is an antiferroelectric. The most complete presentation of results of investigation of phase coexistence in the systems under discussion are presented in [31-34, 37] where $Rb_{1-x}(ND_4)_xD_2AsO_4$ solid solutions are considered as an example (phase diagram is presented in Fig.12). (Before further discussions of this topic we believe that it necessary to point out the resemblance of the $Rb_{1-x}(ND_4)_xD_2AsO_4$ -phase diagram and diagrams depicted in the Fig. 9 for oxides with perovskite structure). The authors of above- mentioned publications have clearly demonstrated the existence of broad regions in the vicinity of the FE-PE and the AFE-PE phase transition lines. These regions are characterized by the coexistence of domains of ordered phases in the paraelectric matrix of the samples. Although, in our opinion this is not the main circumstance

The temperature dependencies of the LA[100] Brillouin backscattering phonon spectra [31] and the Raman vibration modes [32] have been studied in the mixed FE-AFE system of $Rb_{1-x}(ND_4)_xD_2AsO_4$ solid solutions with compositions corresponding to the FE side of the phase diagram. The same investigations for this system of solid solutions with compositions corresponding to the AFE side of diagram have been carried out in [33]. Authors attributed the anomalies of physical properties of the solid solutions with $x = 0.10$ near the temperature $160K$ to the onset of short-range AFE order caused by the freezing-in of $(ND)_4$ reorientations and imply a growth of local structure competition (between FE and AFE ordering). With respect to the opinion of the authors of [31, 32], one can expect that such FE-AFE ordering competition, which can suppress a long-range-order FE transition, is responsible for both phase coexistence and presence of the broad damping peak in the Brillouin backscattering spectrum centered at $T\sim146K$ (Fig.15). The rapid growth of FE ordering near $130K$ is responsible for the Landau-Khalatnikov-like maximum.

Analogous results were also obtained for compositions from the AFE region of the phase diagram for the $Rb_{1-x}(ND_4)_xD_2AsO_4$ solid solutions. Temperature dependencies of the same parameters for the solid solution with x = 0.55 [33] are given in the Fig.15 on the right for illustration purpose. As one can see that in both cases one can safely suggest that the inhomogeneous states of the FE+AFE domains exist in the paraelectric matrix of the $Rb_{1-x}(ND_4)_xD_2AsO_4$ system of solid solutions with compositions both from FE part and AFE part of the phase diagram. These inhomogeneous states exist in wide intervals of thermodynamic variables (such as temperature and composition). It looks like nobody took into account the possibility of these states until the present time. The behaviour of the diffuseness parameters of the phase transition discussed above for $K_{1-x}(NH_4)_xH_2PO_4$ solid solutions has the same nature.

Let us now discuss the situation that takes place in the $Rb_{1-x}(ND_4)_xD_2AsO_4$ series of solid solutions with compositions that characterized by approximately equal stability of the FE and AFE states. Temperature dependencies of parameters of the LA[100] Brillouin backscattering for composition with $x = 0.39$ are presented in the lower left part of the Fig.15. The temperature dependence of the half width of the anti-Stokes line of the Brillouin phonon spectra manifests a wide and intensive maximum in the vicinity of $T = 140K$. As one can see from the phase diagram of $Rb_{1-x}(ND_4)_xD_2AsO_4$ (in Fig.12 ) the line on which the point with this temperature has to be located is absent (we added the said point and the said line ($T_{seg}$) in the diagram). It seems likely that the nature of this anomaly is not clear to the authors of mentioned studies. The satisfactory explanation for it can be found if one uses the concept of coexisting domains of the FE and AFE phases both above and below $T_C$.

In the substances, with the small difference in energies of the FE and AFE phases the domains of the FE and AFE phases coexist in the sample. The dimensions of these domains are determined by a number of factors. The most decisive factor, in our opinion, is the difference of the interplane distances in the crystal lattices of these phases. As noted above, the IDW, which separates the FE and AFE phase domains is coherent. The conjugation of the crystallographic axes along these boundaries is accompanied with elastic stresses, which increase the total energy of the system. Therefore, the smaller is the difference in the crystal lattice parameters for the coexisting FE and AFE phases, the larger IDW area is allowed and the smaller dimensions of FE and AFE phase domain may be.

Hence, it is extremely difficult to identify these states by means of X-ray method or different spectroscopy methods. We as well as other authors encounter this difficulty during investigations of the FE-AFE solid solutions with perovskite structure. Form the physics standpoint such situation may also take place in the system of solid solutions under discussion. In this case the anomaly at the temperature $140K$ in $Rb_{0.61}(ND_4)_{0.39}D_2AsO_4$ (see Fig.15) may be related to the formation of complex FE+AFE domains in the paraelectric matrix of the substance. The dependence of the Brillouin shift on the temperature in this case has to be similar to the one found in [33] and shown in Fig.15.

## 4. CONCLUSION

In our previous publications [4,7,8,10,15] it has been proved that the coexistence of domains of the FE and AFE phases takes place in rather wide range of thermodynamic parameters in the substances with the small difference in free energies of these phases. This state of coexisting domains of the FE and AFE phases leads to a series of peculiarities in behaviour of these systems under external influences (temparutre, electric field, and hydrostatic pressure).

In the present article we demonstrated that the processes of long-duration relaxation of properties are possible the systems with coexisting domains of the FE and AFE phases after these systems were deviated from the state of thermodynamic equilibrium. As a rule, in the case when the peculiarities of behaviour of this kind are manifested, the substances are referred to as belonging to the class of glasses (spin, dipole, etc) and if it had been done once then it was difficult to correct it afterwards because the human thinking is accustomed to stereotypes.

On the other hand, there are some provoking questions about behavioural peculiarities manifested by the substances, which are seemingly reliably referred to as belonging to the class of glasses. In this article, we have considered such state of affairs using some oxides with perovskite structure and KDP-type systems, which have been classified as dipole glasses, as examples. We have demonstrated that that there are peculiarities of behaviour of above-mentioned systems that are explained outside the frames of dipole glass approach, namely, taking into account the coexistence of domains of the FE and AFE phases in these samples.

Effects that we considered are caused by the local decomposition of solid solutions in the vicinity of the boundaries separating domains of coexisting phases. The difference in crystal lattice parameters of coexisting FE and AFE phases and as a concequence the increase of elastic energy along the boundaries separating these domains lead to local decomposition of solid solution. We believe that similar effects might be also possible in the vicinity of domain boundaries between a ferromagnetis and an antiferromagnetic phases in some substances with magnetic ordering.

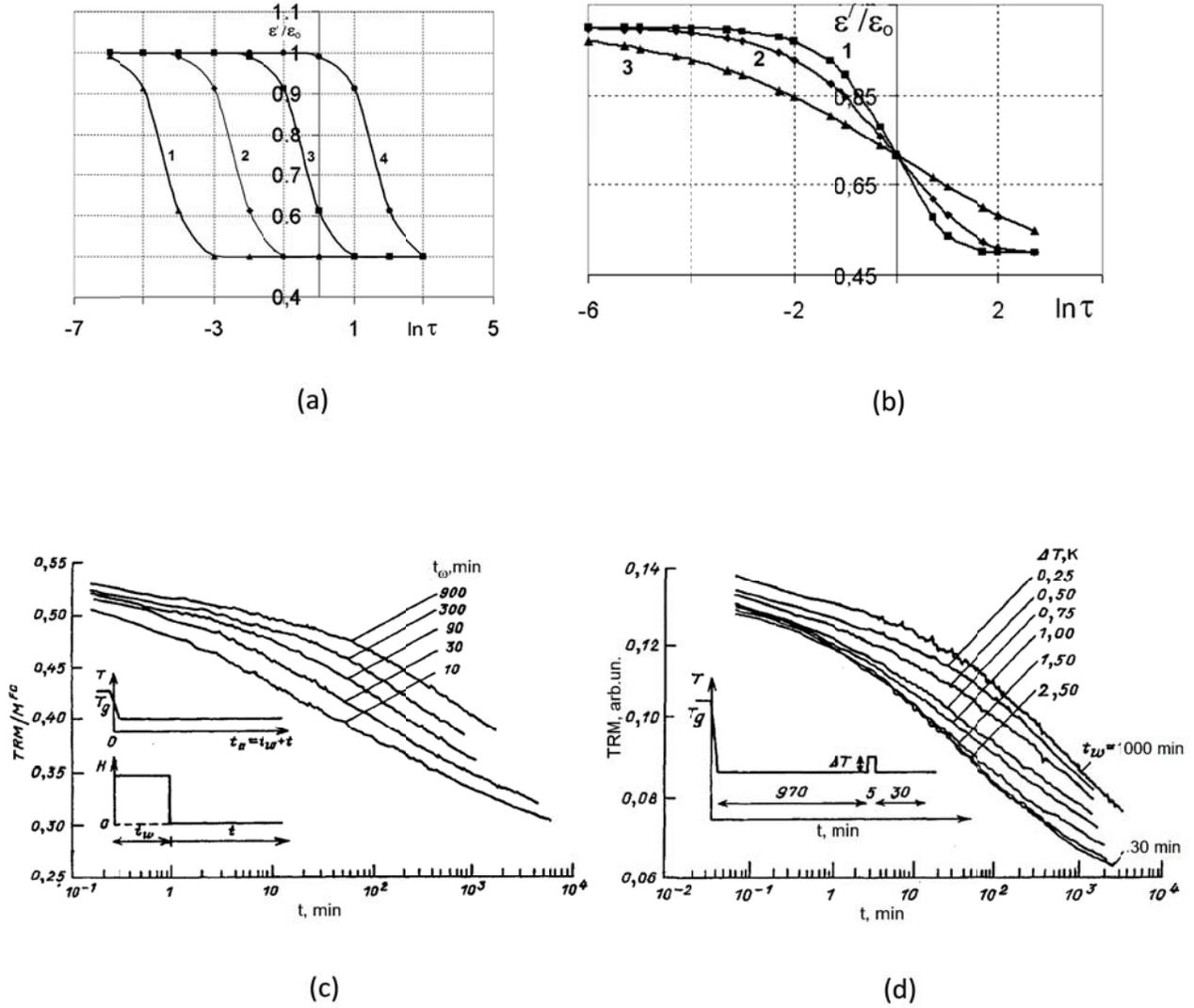

**Figure 1**.

(a) Time dependence of the function $\varepsilon'/\varepsilon_0 = \left[1+\left(1-\exp(-t/\tau)\right)\right]^{-1}$, for different values of τ (1 – τ = 0.0001; 2 – τ = 0.01; 3 – τ = 1.0; 4 – τ = 100.0); (b) Time dependence of the function $\varepsilon'/\varepsilon_0 = \left[1+\left(1-\exp(-t^{1-n}/A)\right)\right]^{-1}$ for different $n$ (1 – n = 0.4, 2 – n = 0.6, 3 – n = 0.8); (c) Magnetization relaxation after the switching-off the magnetic field in experiments on spin-glass aging in zero magnetic field [16]; (d) Relaxation of the magnetization of spin-glass in experiments on aging after temperature heating cycle [16].

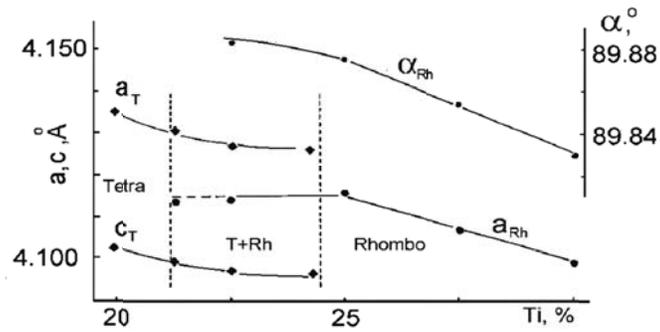

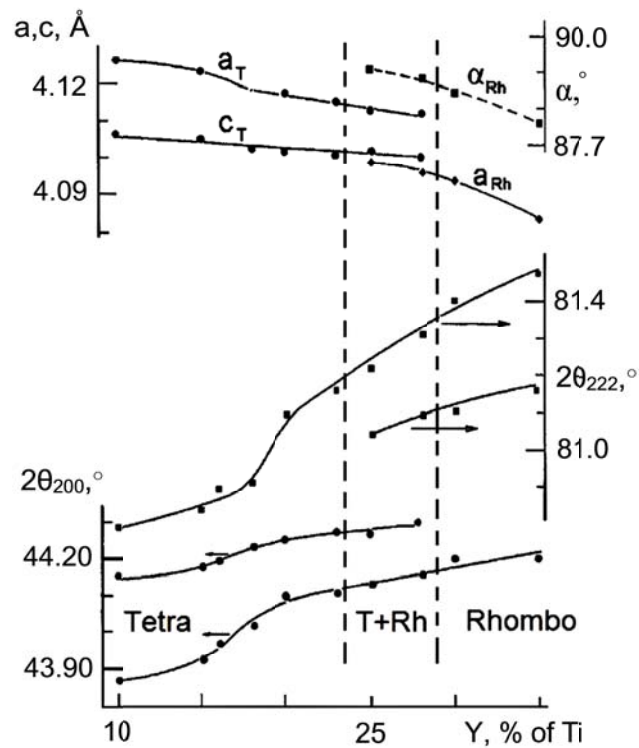

**Figure 2.**

(a) Dependence of lattice parameters on the Ti-content in the PLLZT series of solid solutions 15/100-Y/Y [9, 15]; (b) Dependences of the crystal cell parameters and dependences of the positions of the (200) and (222) X-ray lines (the $2\theta_{200}$ and $2\theta_{222}$ angles) on Ti-content in PLZT solid solutions with 6% of La (6/100-Y/Y) [14].

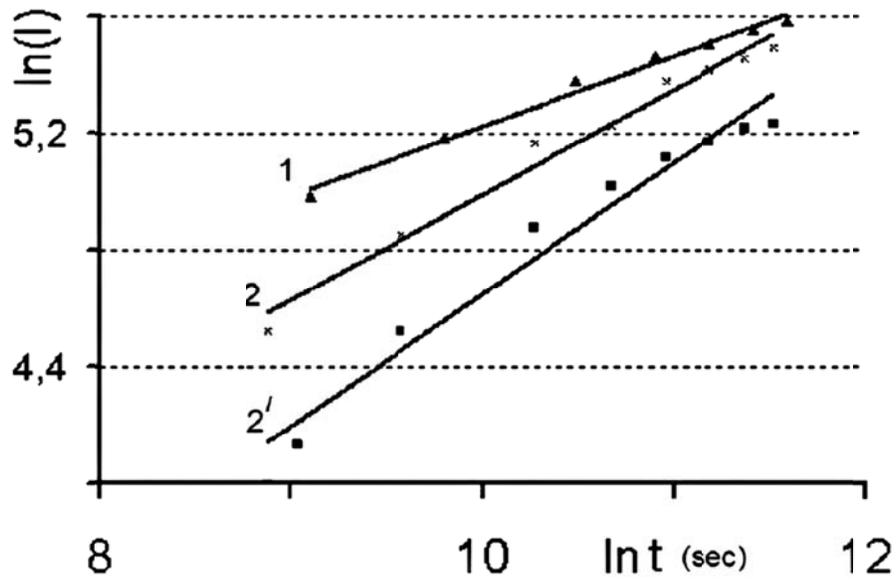

**Figure3.**

Time dependences of the integral intensities of diffusive X-ray lines for 15/77/23 PLLZT (1) and 6/73/27 PLZT (2 and 2$^/$) solid solutions.

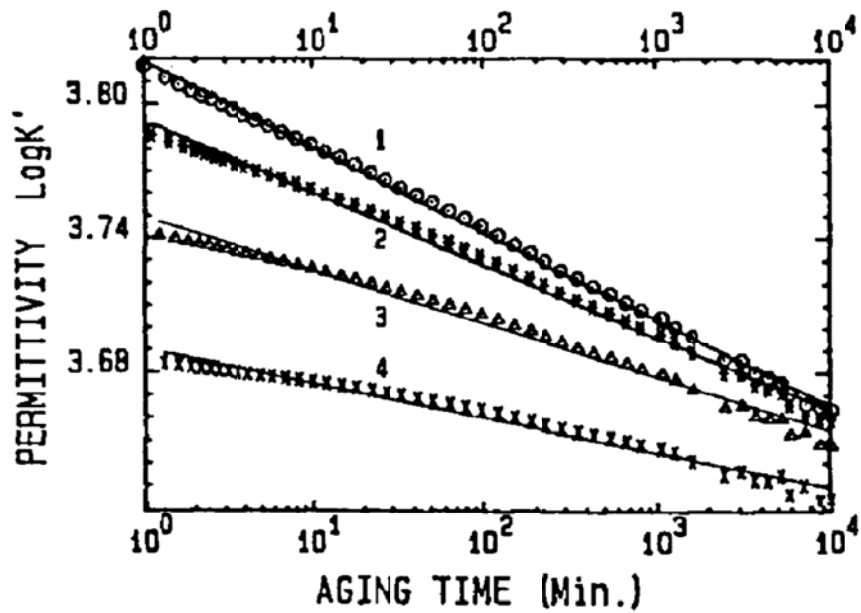

**Figure 4.**

The permittivity as a function of aging time (after quenching from $400^oC$) in 9.5/65/35 PLZT samples [17].

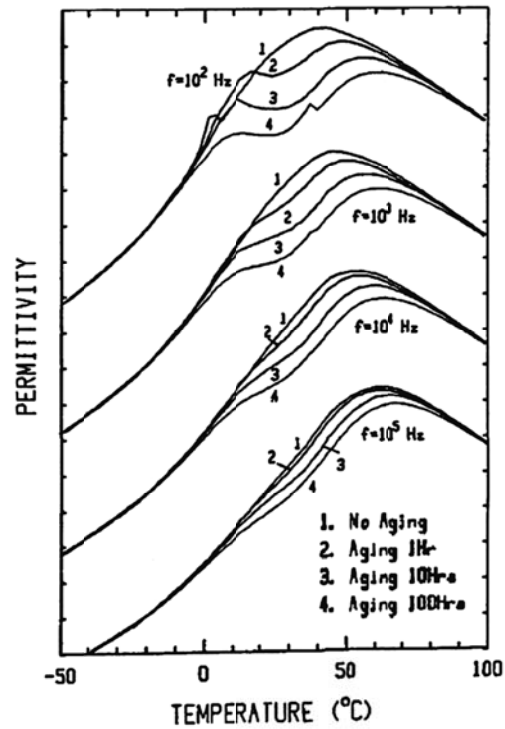

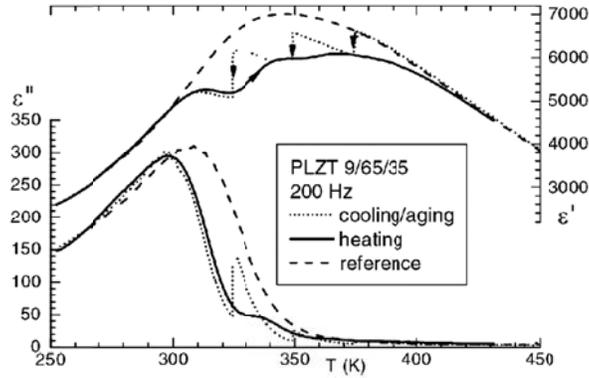

**Figure 5.**

(a) Temperature dependences of dielectric constant for different aging time (after quenching from $400^{o}C$ and aging at the temperature ~ $23^{o}C$) in 9.5/65/35 PLZT samples [17]; (b) Dielectric permittivity of 9/65/35 PLZT samples at 200 Hz during multiple aging stages (24 *h* each) and subsequent heating curves with memory, compared with the reference curve measured on continuous cooling [18].

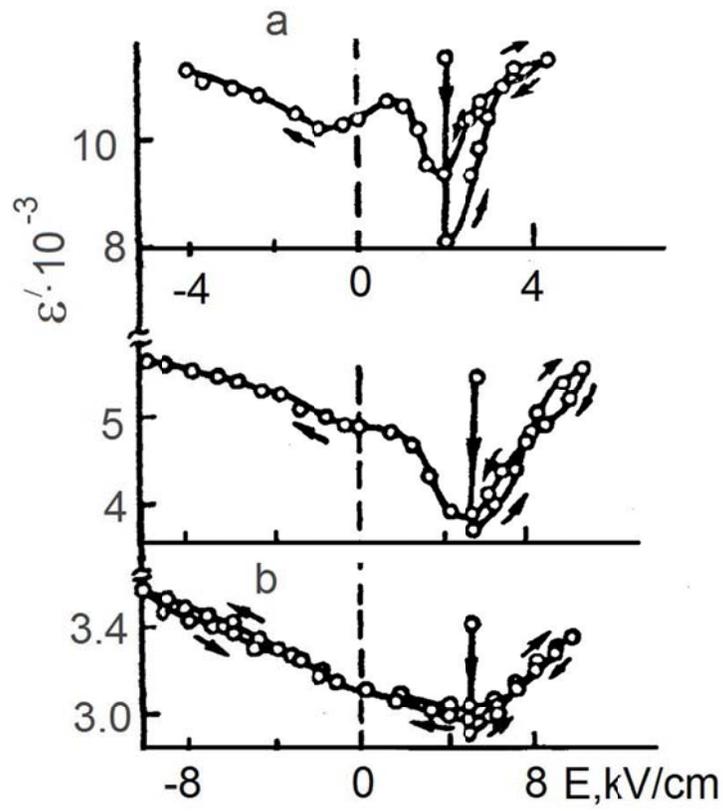

**Figure 6**.

Dependence of dielectric constant on DC bias after aging during 20 h. The compositions of solid solutions: a – PLZT 8/65/35, $T_{age}$ = 57°C; b – PLZT 11/65/35, $T_{age}$ = 22°C; c – PLZT 13.5/65/35, $T_{age}$ = 22 °C [19].

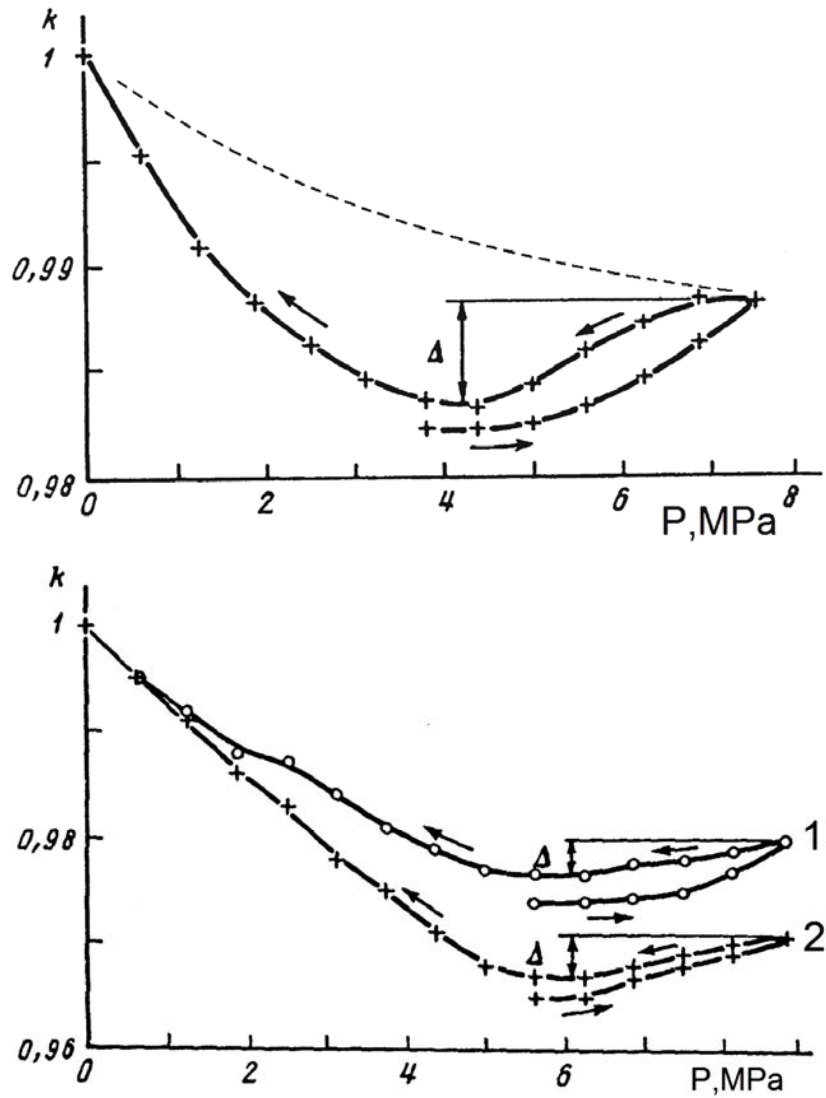

**Figure 7.**

On the top: Dielectric constant vs axial stress for 8/65/35 PLZT after 20 h aging [20] at the temperature $T_{age}$ = 27 °C and pressure $P_{mech}$ = 3.8 MPa. Dashed line (added by the authors of the present paper) presents the dependence without aging.
On the bottom: Dielectric constant vs axial stress for 8/65/35 PLZT after aging in the presence of the electric field E = 1850 V/cm for 48 h (1) and 120 h (2) [20].

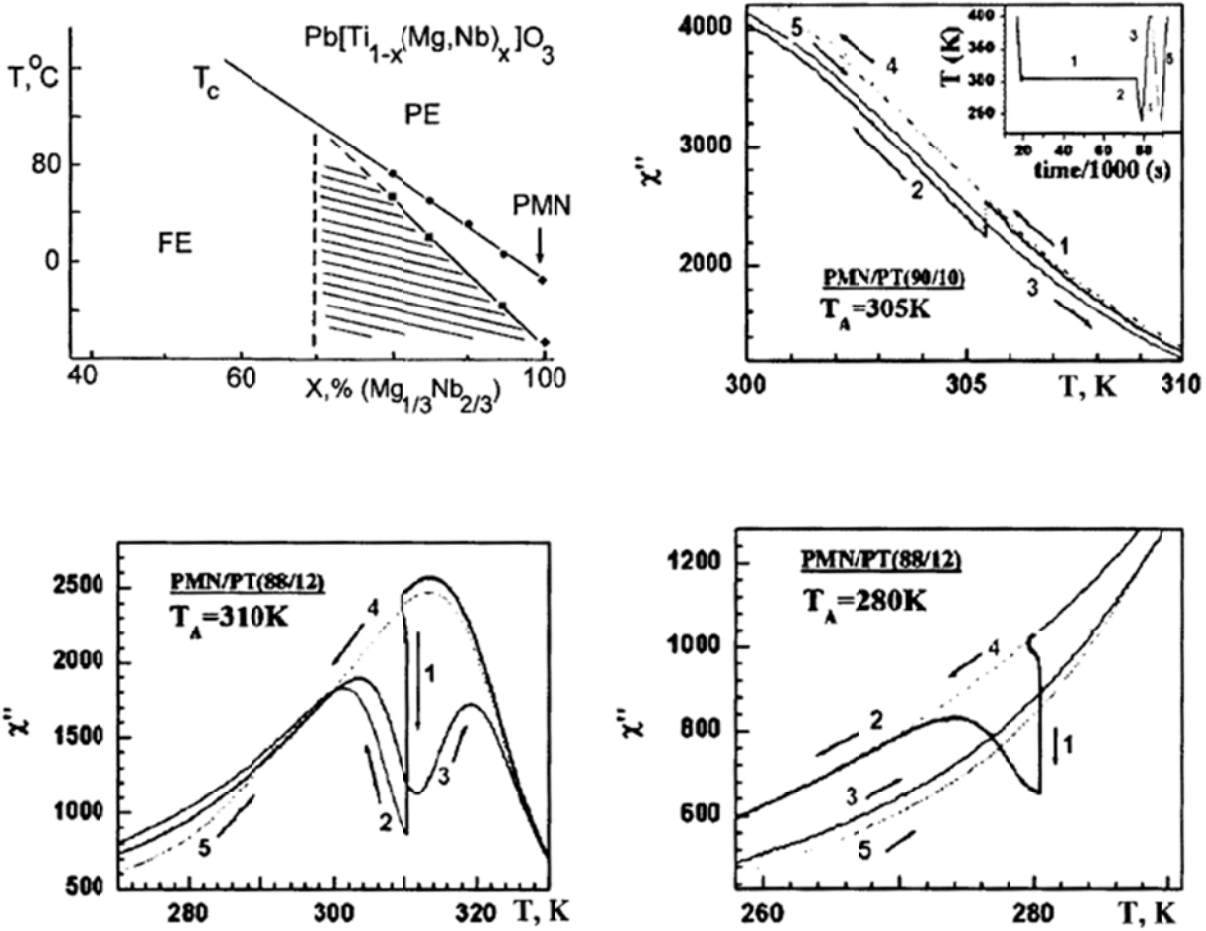

**Figure 8.**

Phase diagram of the (1-x)Pb[(Mg$_{1/3}$Nb$_{2/3}$)O$_3$ - xPbTiO$_3$ system of solid solutions (constructed based on data from [38-45]) and dielectric memory effect in these solid solutions [22].
In a typical aging memory experiment (see insert for temperature history profile) the sample is cooled to $T_A$ (curve 1). After aging at $T_A$, the sample is cooled to a lower excursion temperature $T_{EX}$ (curve 2) and then immediately reheated past $T_A$ (curve 3). For comparison, reference starting curves (4 and 5) are also taken at the same sweep rates without stopping at $T_A$.

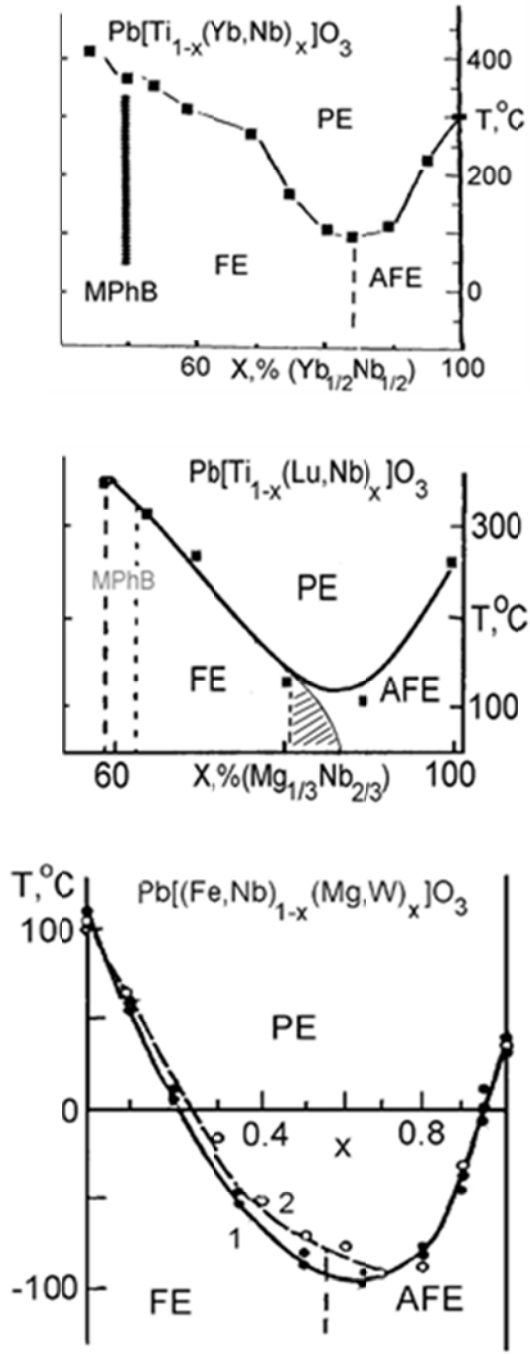

**Figure 9.**

Phase diagrams of solid solutions with perovskite structure obtained by the substitution of titanium by the complex: $(Yb_{1/2}Nb_{1/2})$ [24], $(Lu_{1/2}Nb_{1/2})$ [23], and $Pb(Mg_{1/2}W_{1/2})O_3$ [25, 26].

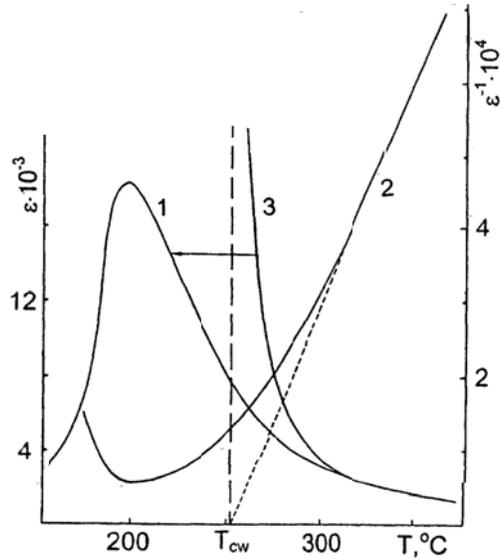

**Figure 10.**

Temperature dependences of ε (1) and 1/ε (2) for ferroelectric near the FE-AFE-PE triple point (with DPT) and those calculated from Curie-Weiss low ε(*T*) dependence (3). The arrow shows the shift of ε(*T*) dependence under effect of FE-nonactive impurity.

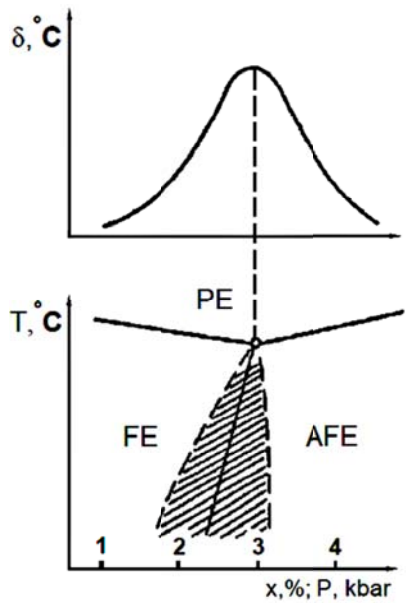

**Figure 11.**

Generalised diagram of phase states (at the bottom) and the dependence of diffuseness parameter on external factors (at the top) near FE-AFE-PE triple point.

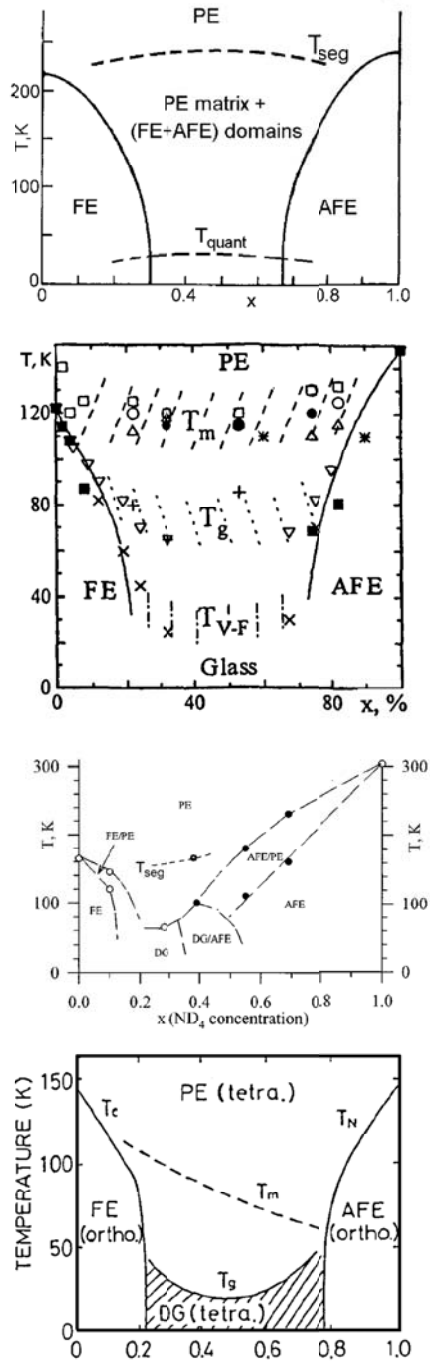

**Figure 12.**

Phase diagrams of solid solutions with order-disorder phase transitions. From top to bottom: model diagram, diagram for $K_{1-x}(NH_4)_xH_2PO_4$ [30], diagram for $Rb_{1-x}(NH_4)_xD_2AsO_4$ [33], the line $T_{seg}$ is added by authors of the present paper based on results of [31-34], and diagram for $Rb_{1-x}(NH_4)_xH_2PO_4$ [35].

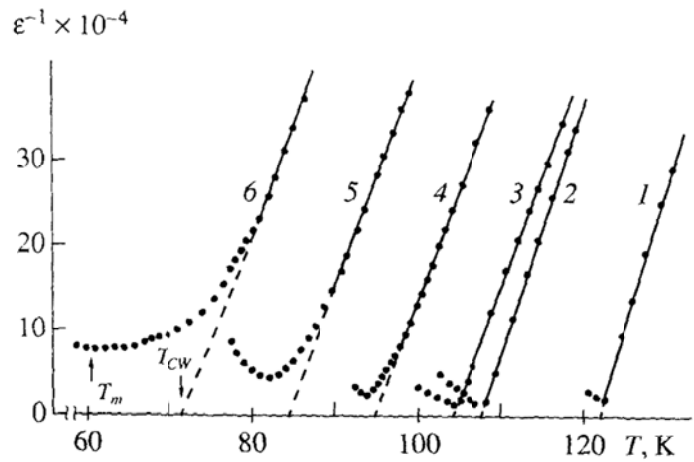

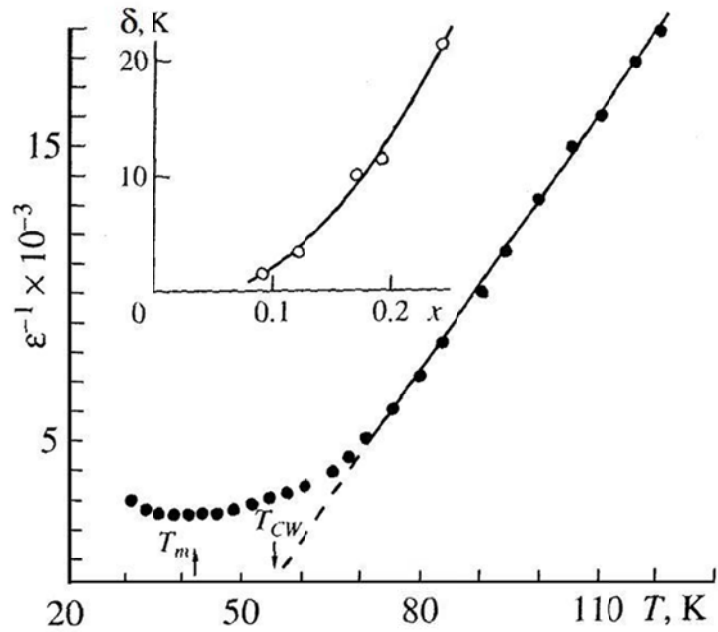

**Figure 13.**

On the top: Temperature dependencies of $\varepsilon^{-1}$ for the crystals of the $K_{1-x}(NH_4)_xH_2PO_4$ series [36]. Compositions x = 1 – x = 0; 2 – x = 0.04; 3 – x = 0.05; 4 – x = 0.09; 5 – x = 0.12; 6 – x = 0.19.
On the bottom: Temperature dependencies of $\varepsilon^{-1}$ for the $K_{0.76}(NH_4)_{0.24}H_2PO_4$ crystal. Dependence of the diffuseness parameter on the composition is given in the insert [36].

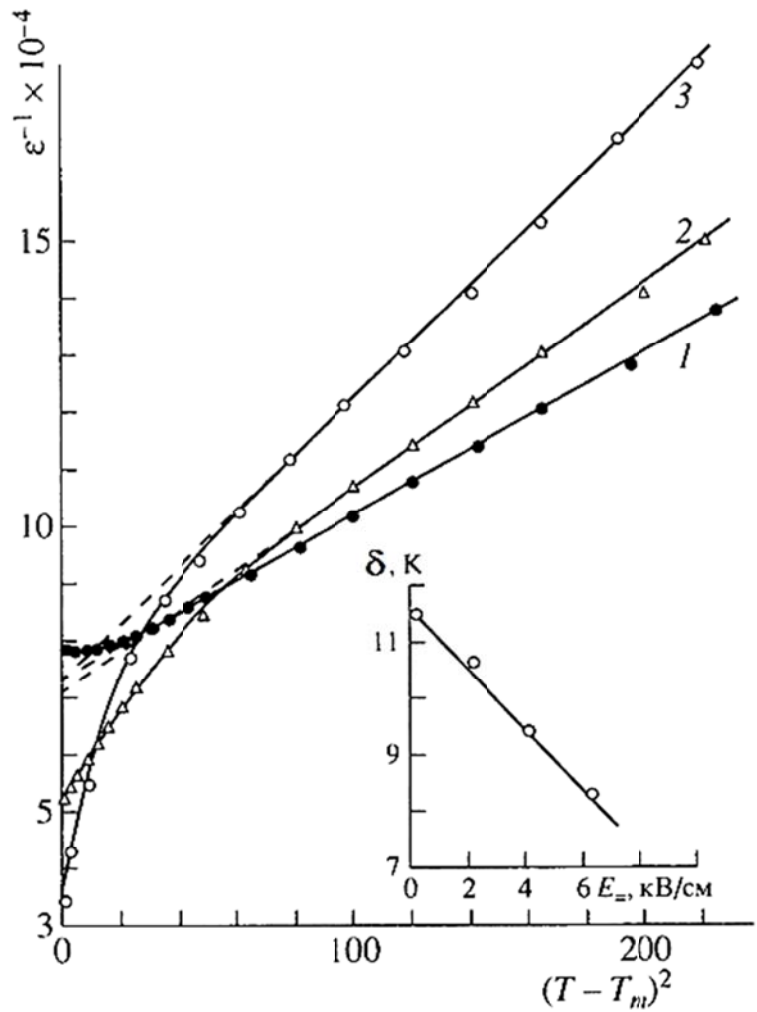

**Figure 14.**

Temperature dependencies of $\varepsilon^{-1}$ for the solid solution $K_{1-x}(NH_4)_xH_2PO_4$ with x = 0.19 in external electric field $E$. Electric field ($kV/mm$): 1 – 0; 2 – 4; 3 – 6 [36]. Dependence of the diffuseness parameter on the electric field $E$ is given in the insert.

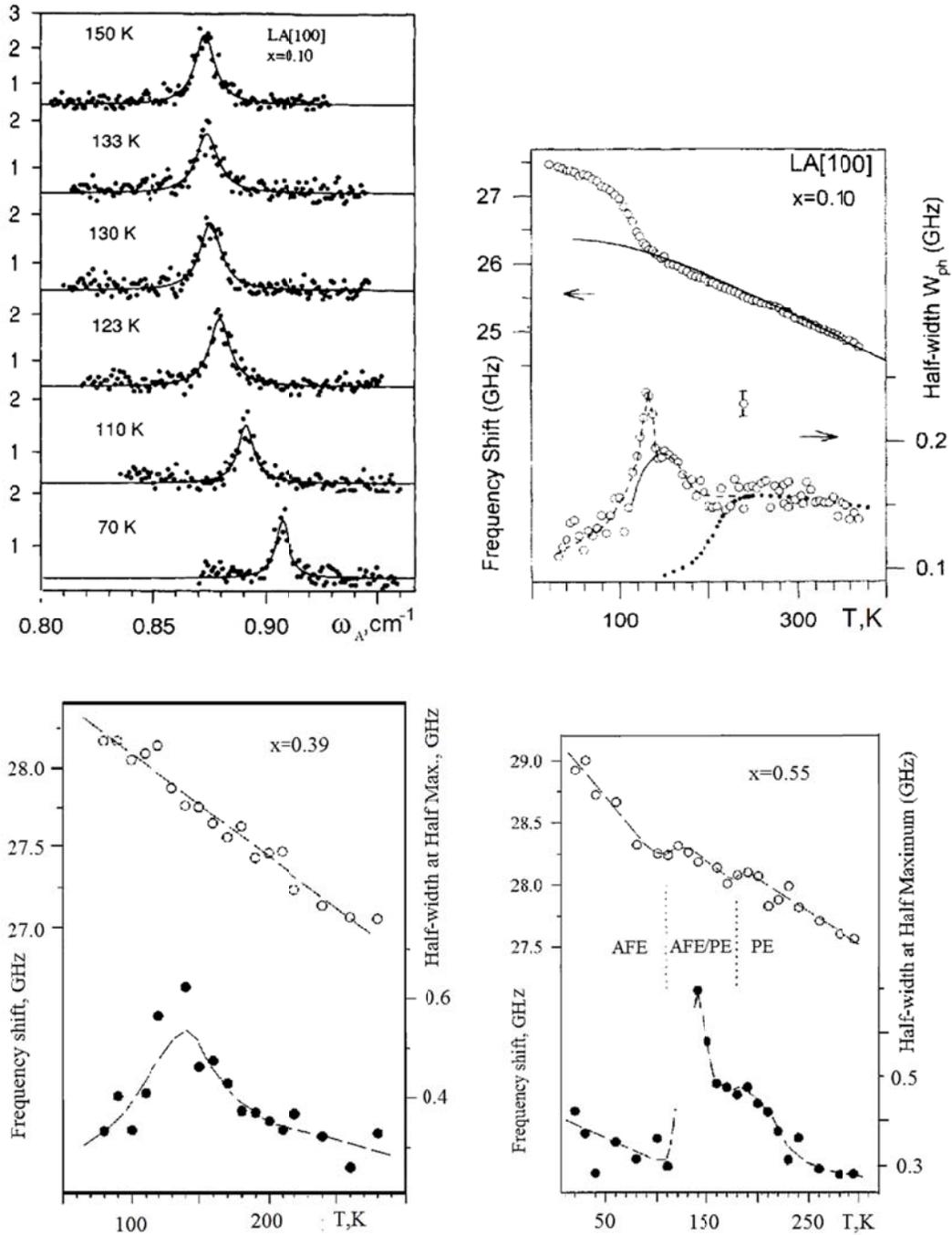

**Figure 15.**

Anti-Stockes component of LA [100] Brillouin frequency shift for temperatures around the maximum value of half-width in $Rb_{1-x}(ND_4)_xD_2A_sO_4$ mixed crystals (x = 0.10) [31] (upper left). The Brillouin shift (open circle) and the half width (solid circle) vs temperature of the LA[100] phonons for x = 0.10 [31] (upper right);  x = 0.39 [33] (lower left);  and  x = 0.55 [33] (lower right).